# Application of avalanche photodiodes as a readout for scintillator tile-fiber systems

C. Cheshkov [a], G. Georgiev [b], E. Gouchtchine [c], L. Litov [a], I. Mandjoukov [a], V. Spassov [d]

[a] *Faculty of Physics, Sofia University, 5 James Bourchier Blvd., 1164 Sofia, Bulgaria*

[b] *Institute for Nuclear Research and Nuclear Energy, Bulgarian Academy of Sciences, 72 Tzarigradsko Shousse Blvd., 1784 Sofia, Bulgaria*

[c] *Moscow Institute for Nuclear Research, 60th Anniversary Prospekt, RU-117 312 Moskva, Russia*

[d] *"Inter Q" Ltd., 1 Kukush str., 1309 Sofia, Bulgaria*

**Abstract**

The application of reach-through avalanche photodiodes (R'APD) as a photodetector for scintillator tiles has been investigated. The light collected by WLS fibers ($0.84mm$ and $1mm$ diameter) embedded in the scintillator has been transmited to the $0.5mm^2$ active surface of APD by clear optical fibers and optical connectors. A low noise charge sensitive preamplifier ($\approx 400e^-$ equivalent noise charge) has been used to gain the photodiode signal. Various configurations of tile-fibre systems, suitable for CMS and LHCb experiments at LHC have been studied using cosmic muons and muon beam at SPS at CERN. In order to optimize the performance of APD, measurments in the temperature range from $-10^oC$ to $+25^oC$ have been done. The $MIP$ detection efficiency and $e^-/MIP$ separation have been estimated in order to determine applicability of the readout for LHCb preshower.

## 1 Introduction

Recently intensive R&D on scintilator tile-fiber readouts is being carried out in order to satisfy the needs of calorimeters in new LHC experiments [1], [2]. The specific requirements for these types of detectors are:

- Operation in magnetic field up to $4Tesla$.

- Large linear dynamic range - $10^5$ for the detector-preamplifier couple.
- Long lifetime of the photodetectors - 5 to 10 years of operation at high luminosity.
- Radiation hardness - up to $2Mrad$ integrated dose.
- Small size - because of very large number of channels in use.
- Sufficient $Signal/Noise$ ratio - to measure the signal from a minimum ionising particle ($MIP$).
- Capability of measuring the signal generated by a radioactive source as a DC current to a precision of 1%.
- Reasonable price.

Two main types of photodetectors satisfy the above requirements: hybrid photodiodes (HPD) and avalanche photodiodes (APD). The advantages of avalanche photodiodes over the other types of photodetectors are:

- Insensitivity to magnetic field.
- Linear dynamic range of $10^6$.
- Fast response ($< 1ns$).
- High quantum efficiency in the range from $200nm$ to $1100nm$.
- Small size ($10 \times 10 \times 5mm^3$).
- Low price of APD and preamplifier.

Possible disadvantages of the photodiodes are low internal gain ($50 - 300$) compared to the PMT and HPD one and relativelly high excess noise factor. The present paper is devoted to the investigation of the applicability of APD's as a readout for scintilator tile-fiber systems. The choice of the scintilator tiles design has been determined by the requirements for two of the LHC experiments under preparation - CMS and LHCb. The main goal of our research was to achieve $Signal/Noise$ ratio enough to measure the $MIP$ signal from scintilator tile-fiber and good $e^-/MIP$ separation needed for the preshower detector of the LHCb experiment. For this purpose, various designs of the scintilator tile-fiber system have been developed (section 2). The APD characteristics, electronics and calibration of the systems under investigation are presented in sections 3, 4 and 5 correspondently. The results from measurements performed with cosmic muons and muon beams at SPS accelerator are reported in section 6.
It is shown that cooling of the photodiodes reduces dark current and increases gain hence allowing us to achieve by times higher $Signal/Noise$ ratio and to reach our goals.

## 2 Scintillator tile - fibre system

Two different designs of scintillator tile - fibre system have been studied (fig.1). The first one is proposed for CMS HCAL [3] while the second one is under investigation for the preshower detector in LHCb experiment [4]. Three different types of WLS fibers - green Kuraray Y11 /0.84$mm$ diameter/, green Bicron 91A /1$mm$/ and red Bicron 99-172 /1$mm$/ have been used. In order to estimate the scintilators light output, cosmic muons measurments using PMT FEU85 have been performed. The WLS fibers were directly coupled to the PMT window. The signal from PMT was read out by charge-sensitive ADC LeCroy 2249W with 250$fC/channel$ sensitivity using a 160$ns$ gate triggered by two scintillator counters. Calibration of PMT was done by fitting the single photoelectron distribution, produced by 15$ns$ light pulses of blue LED [5]. The average number of photoelectrons induced by cosmic muons are presented in Table 1. Taking into account the PMT photocathode quantum efficiency at 500$nm$ emission peak of green WLS fibers ($\approx 6\%$) [6] one can estimate $\approx 180$ photons from $4 \times 4 \times 1cm^3$ scintillator with 8 coils of fiber. Coupling of WLS fiber to a clear fiber with the help of optical connector leads to the reduction of the light output at the level of $20\% - 35\%$. In order to improve the light collection we have matted the scintillator surface on which the WLS fibers were placed. This gaves a 20% increase of the light yield.

Table 1
Average number of photoelectrons induced by cosmic muons.

| Tile-fiber configuration | Light yield, ph.e. |
|---|---|
| $22 \times 22 \times 0.4cm^3$ scintillator with 1 coil of Bicron 91A fiber | 4.0 |
| $22 \times 22 \times 0.4cm^3$ scintillator with 3 coils of Bicron 91A fiber | 7.8 |
| $4 \times 4 \times 1cm^3$ scintillator with 8 coils of Bicron 91A fiber | 11.1 |

## 3 Avalanche Photodiodes

The avalanche photodiodes under investigation are manufactured by "InterQ" Ltd. using planar technology on a $p-$type high resistivity silicon with a resistivity of $2k\Omega cm$ [7],[8]. They are $n^+ - p - \pi - p^+$ type (R'APD).The multiplication region of the diodes is produced by ion implantation and two-stage diffusion of boron and phosphorus (fig.2). The active surface of the detectors is $0.5mm^2$. Some characteristics of the diodes are presented in Table 2.

Table 2
Avalanche photodiode characteristics at room temperature.

| | |
|---|---|
| Thickness [$\mu m$] | 140 |
| Active dia. [mm] | 0.8 |
| Capacity [$pF$] at 100$V$ | 0.92 |
| Breakdown voltage [$V$] | $250 - 380$ |
| Dark current [$nA$] at 100$V$ | $< 1$ |

## 4 Electronics

The schematic view of the specialy designed for our investigation electronics is presented on fig.3. The chain consists of a charge-sensitive preamplifier, CR-RC shaper and current feedback amplifier, which can drive a back terminated 50Ω line. A JFET SST309 transistor with high forward transconductance (determined by the very short shaping time - 25 ns) about 15 mS per 10 mA drain current and a relatively low input capacity - 6 pF has been used. The noises of the shaping amplifier become significant (the noise bandwidth is 10 MHz) due to the short shaping time. For this reason, a microwave bipolar transistors with low $R_{BB'}$ have been used. The thermostability of the preamplifier was achivied via negative feedback.
Main parameters of the preamplifier are:

- Equivalent charge noise $\approx 400e^-$ at $1pF$ detector capacity (fig.4).
- Sensitivity of the preamplifier is $160mV/10^6e^-$. (This coefficient depends on value of capacitor $C1$).
- $25ns$ shaping time.
- Full width of output signal $\approx 150ns$.
- Maximum output voltage about $3.5V$.

## 5 Calibration of the readout

The calibration of the readout has been done using $15ns$ long blue light pulses emited by LED. The light has been splited up between APD and PMT readouts throught a special optical connector. The signals from both readouts have been registrated by qADC with $160ns$ gate. The measurements have been done at different light intensities, reverse bias voltages and temperatures . The summary results are presented below:

- The normalized internal gain of APD as a function of applied bias voltage is shown on fig.5. The gain at $60V$ and $22^oC$ is about 6. Cooling of the photodiodes leeds to breakbown voltage decrease and to a valuable gain increase mainly at voltages near to the breakdown.

- The APD and preamplifier noise in terms of ADC channels vs. applied bias voltage is ploted on fig.6. At low temperatures the noise remains low up to particular voltage near to the breakdown and then increases drastically. On the other hand at high temperatures, noise is higher and increases smoothly. The equivalent noise charge can be estimated taking into account that the preamplifier gain is $68e^-$ per ADC channel. The small increase of noise at low bias voltages is due to larger capacity of photodiodes which is a result of the absence of full depletion of charges in APD at these voltages.
- Fig.7 shows the Signal/Noise ratio as a function of the bias voltage. The cooling of APD allow us to increase drastically the Signal/Noise ratio. The maximum of this ratio can be achivied at bias voltages near the breakdown where the signal is already large while the noise is still low. Further increase of the voltage results in abrupt increase of the noise.

It's clear that even for low intensity ligth sources, cooled APD gives a considerable S/N ratio. The large increase of the gain of APD at low temperatures is a result from fact that the multiplication process in the APD is affected by the temperature. This happens since electrons lose energy to the phonons, whose energy density increases with the temperature, and at lower temperatures it takes shorter for the electrons to reach the energy required for impact ionization.

## 6 Measurments with muons

We have performed series of measurments of the scintilator tile-fiber readout using cosmic muon flux and muon beam at SPS at CERN. The tests have been made with PMT and APD readouts in a wide temperature range. We have obtained $1.5-2$ times higher efficiency of Bicron 99-172 fiber over Bicron 91A fiber and $1.25-1.5$ times over Kuraray Y11 fiber with APD readout because of the higher photodiode quantum efficiency at longer wavelengths of incoming light. The results obtained with Bicron 99-172 fiber are presented on fig.8 and fig.9. Due to the long right-handed tail in the muon signal distribution, in what follows the $S/N$ ratio is defined as a ratio of the signal distribution maximum $MPV$ (Most Probable Value) [9] to the $\sigma$ of pedestal. The tail is determined by a low light yield from the scintillator and by the so called excess noise factor (F) of the photodiodes. Taking into account that F rises linearly with the bias voltage applied, we have to find the operational voltage at which the $S/N$ ratio is enough good, keeping the RMS of the signal distribution as low as possible. We have found that the compromise is reached at $U \approx 110V$ (see fig.10) for the photodiode used.

We have calculated the $MIP$ detection efficiency (the ratio of the number of events in the signal distribution above $2\sigma$ from the pedestal to the number of all events in the signal distribution) for different bias voltages (fig.12). Another important issue is the ability to separate clearly the signals from muons ($MIP$) and electrons. For example, for the $LHCb$ preshower detector the signal is considered as a $MIP$ when it is by five times lower than the muon $MPV$. Otherwise it is treated as an electron signal. A separation ratio (the ratio of the number of the events in signal distribution below $5MPV$ to the number of all events in the signal distribution) as a function of the bias voltage is presented in fig.13. For the bias voltage of 110 V we have satisfactory levels for the detection efficiency $\approx 85\%$ and a separation ratio of $\approx 95\%$.

We also have performed tests with different gates (fig.11) in order to investigate the behavior of the readout system in various readout schemes. The best result for the $S/N$ ratio is when the ADC gate is near to full width of the signal ($160ns$). At narrower gates (for example $50ns$) $S/N$ decreases by 25%. At wider gates $S/N$ falls down rapidly because much more noise is integrated.

## 7 Conclusions

Avalanche photodiodes were tested as a scintilator tile-fiber readout for CMS HCAL and LHCb ECAL preshower. Various types of fibres were investigated and Bicron 99-172 one was determined to be the most efficient. A low noise charge sensitive preamplifier with $\approx 400e^-$ equivalent charge noise was designed to gain signal from photodiodes. For LHCb ECAL preshower scintilator a $Signal/Noise$ ratio of 6.5 and for CMS HCAL scintillator $Signal/Noise$ ratio of 1.3 were achieved cooling the APDs down to $-10^oC$. A satisfactory $MIP$ detection efficiency of $\approx 85\%$ and an excellent $e^-/MIP$ separation of $\approx 95\%$ for preshower scintilator were reached.

## 8 Acknowledgements

We would like to express our gratitude to Dr. Y. Musienko for the useful discussions and consultations and to P.Slavchev ("Inter Q" Ltd.) for developing and manufacturing the cooling device.

## List of Figures

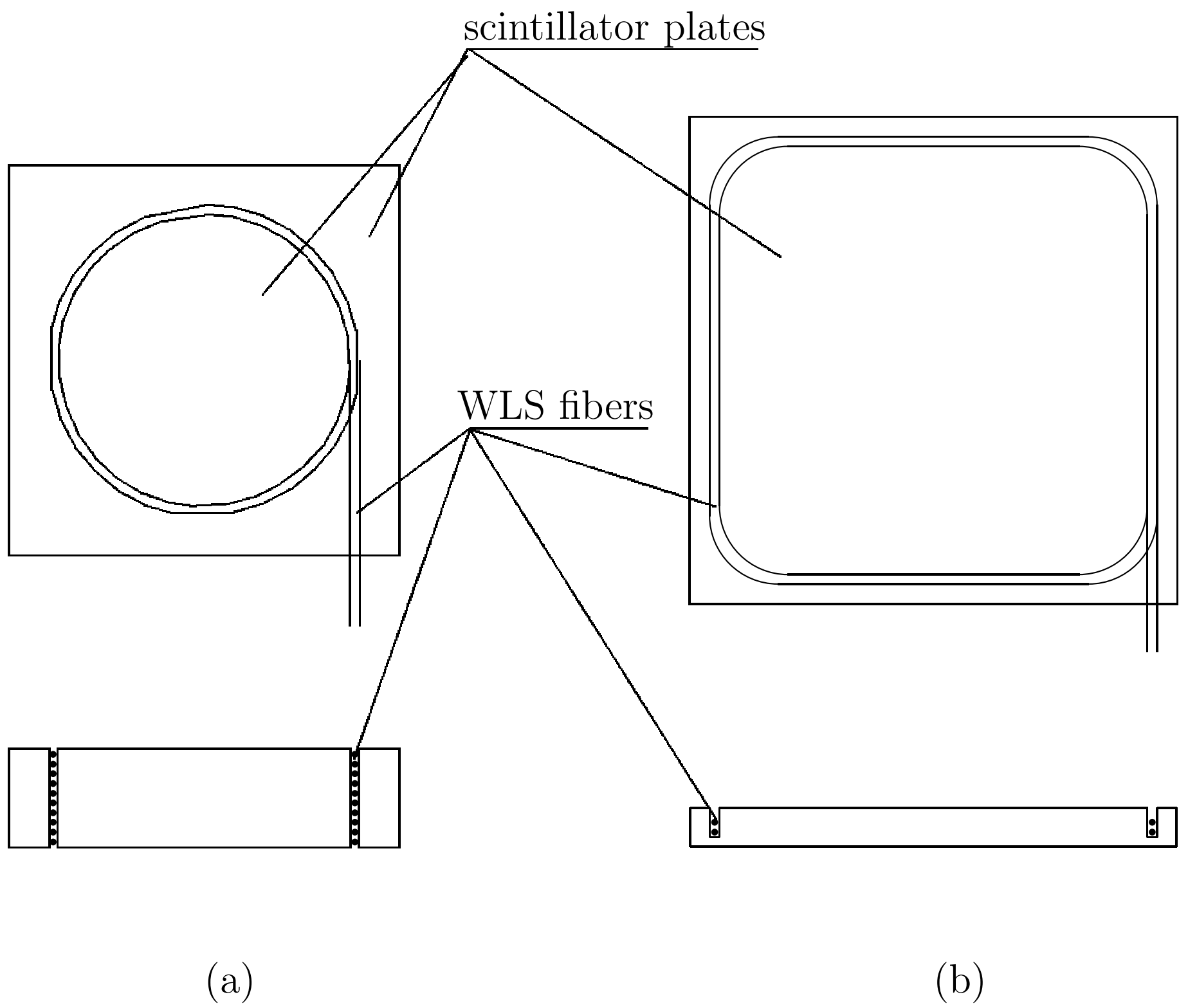

Fig. 1. Two designs of the scintillator tile - fibre system. (a) scintillator size $4 \times 4 \times 1 cm^3$; (b) scintillator size $22 \times 22 \times 0.4 cm^3$.

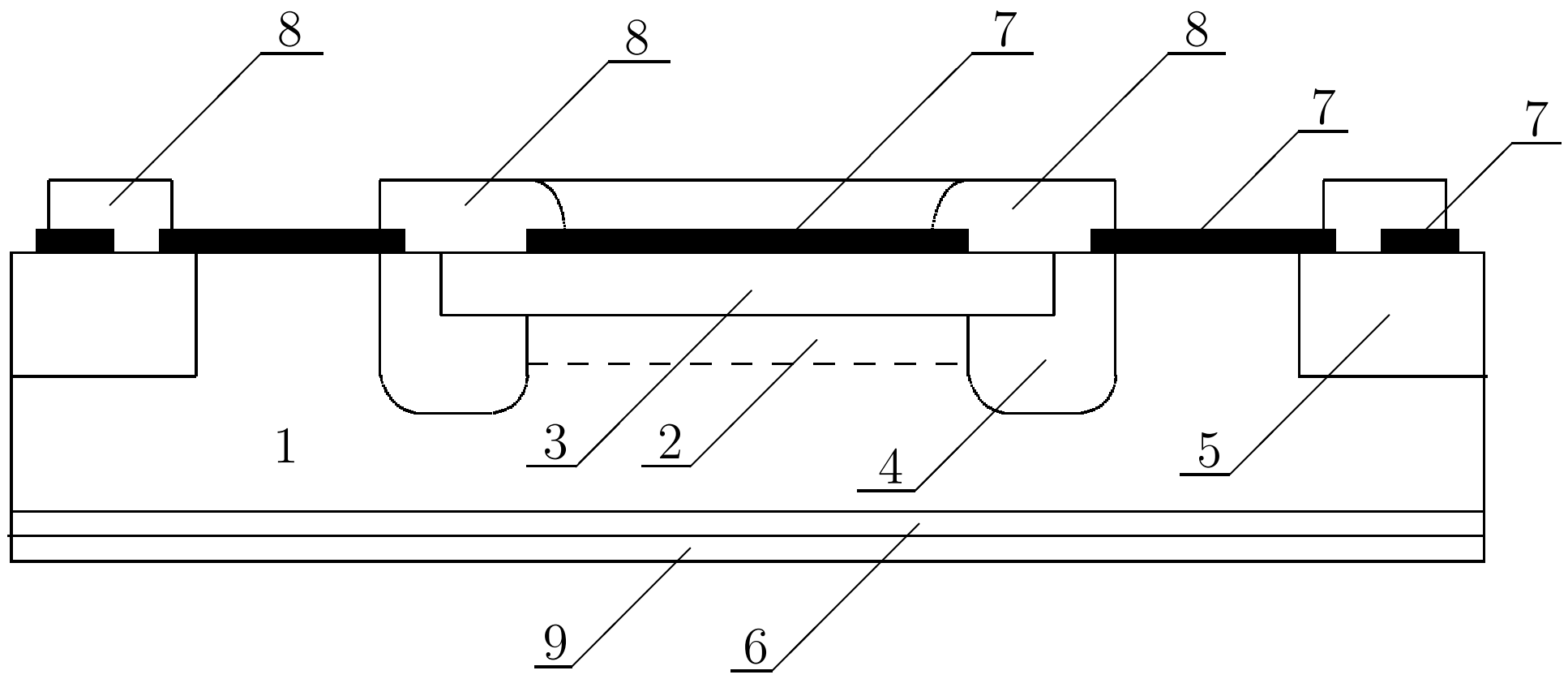

Fig. 2. A transverse view of the avalanche photodiode. (1) high resistivity $p-Si$; (2) $p-$region; (3) $n^+-$region; (4) $n-$region; (5) $p^+-$region; (6) $p-$region; (7) dielectric cover; (8),(9) metal layers.

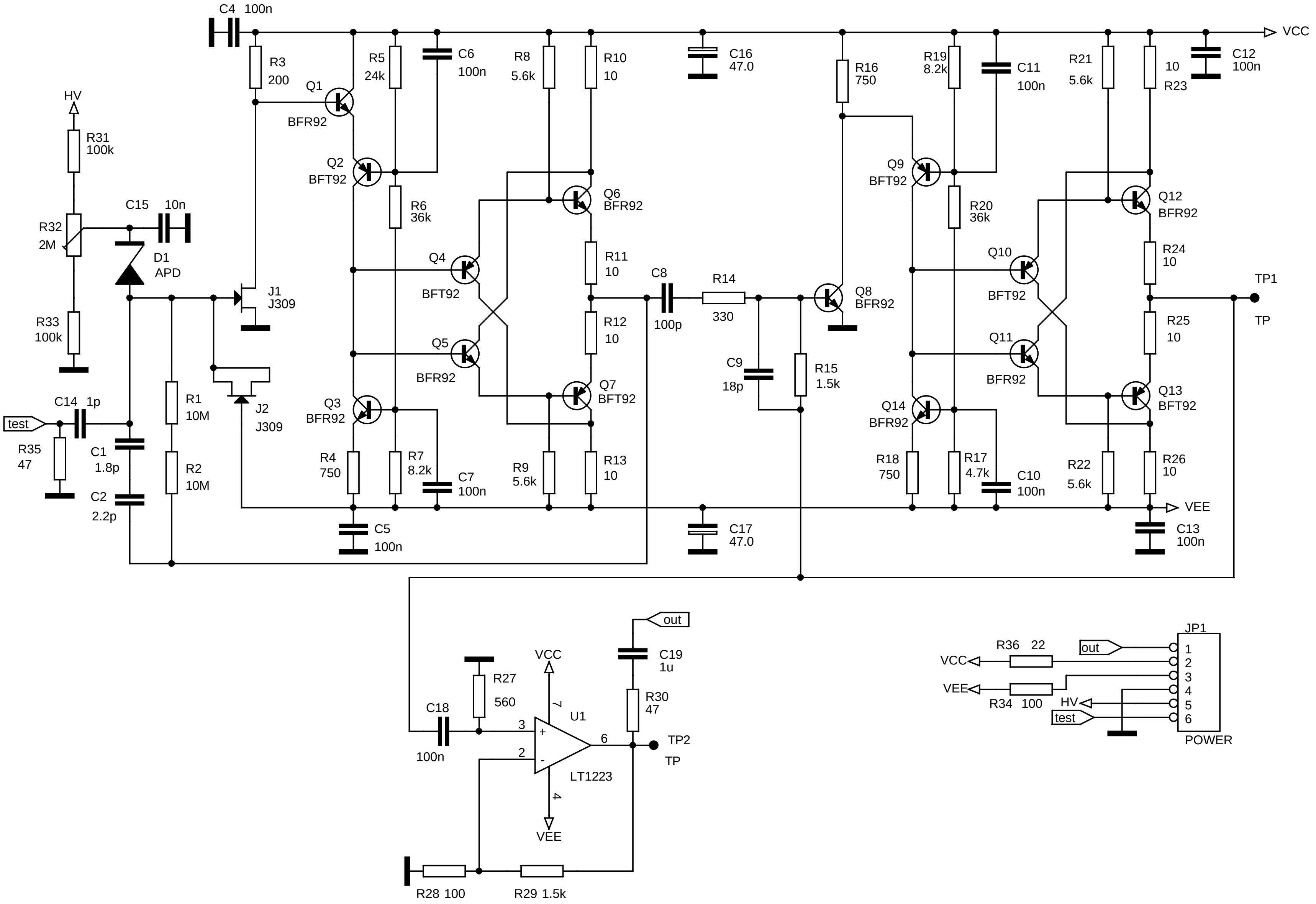

Fig. 3. Schematic view of the electronics.

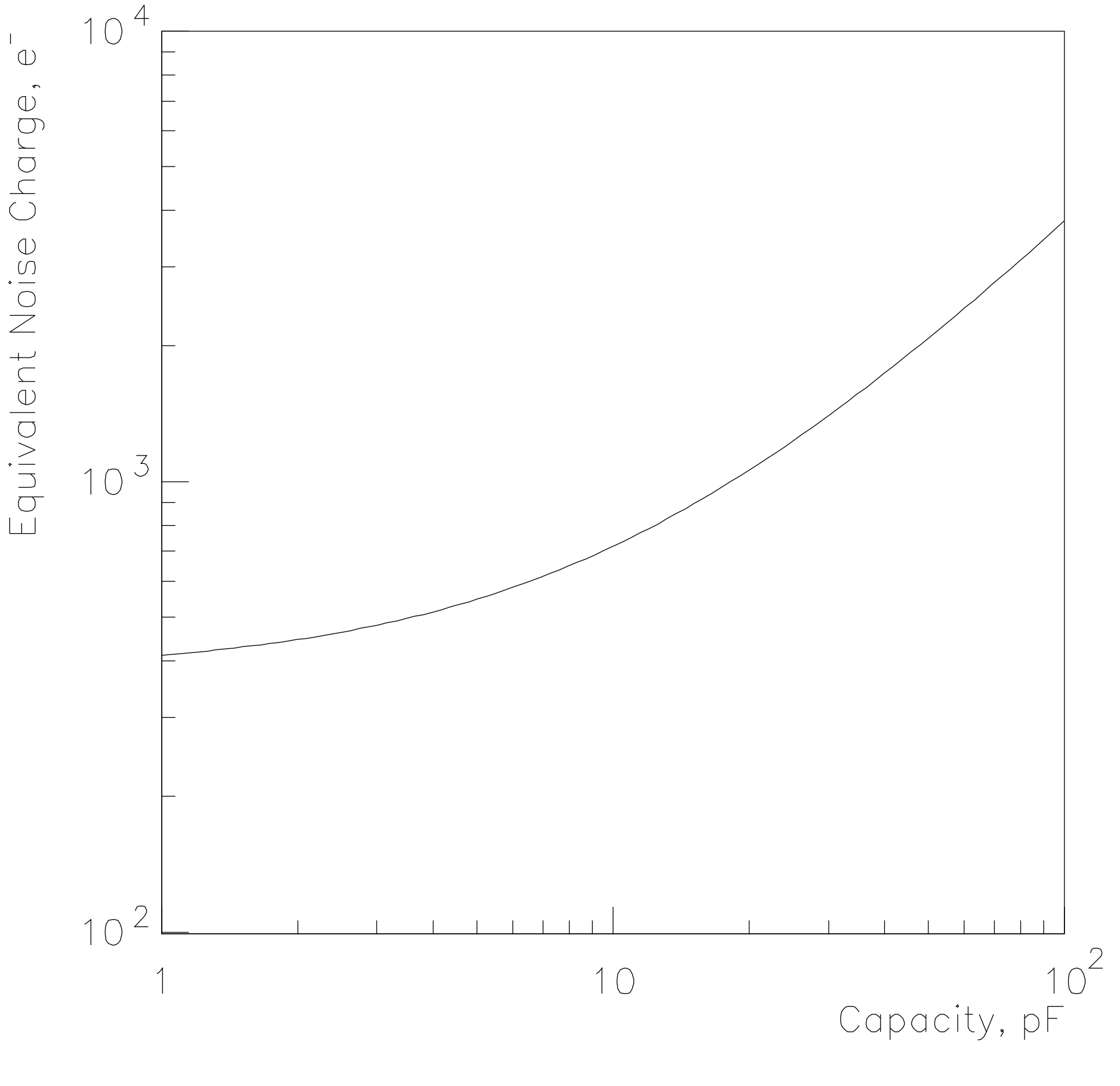

Fig. 4. The equivalent noise charge of the preamplifier as a function of the detector capacity.

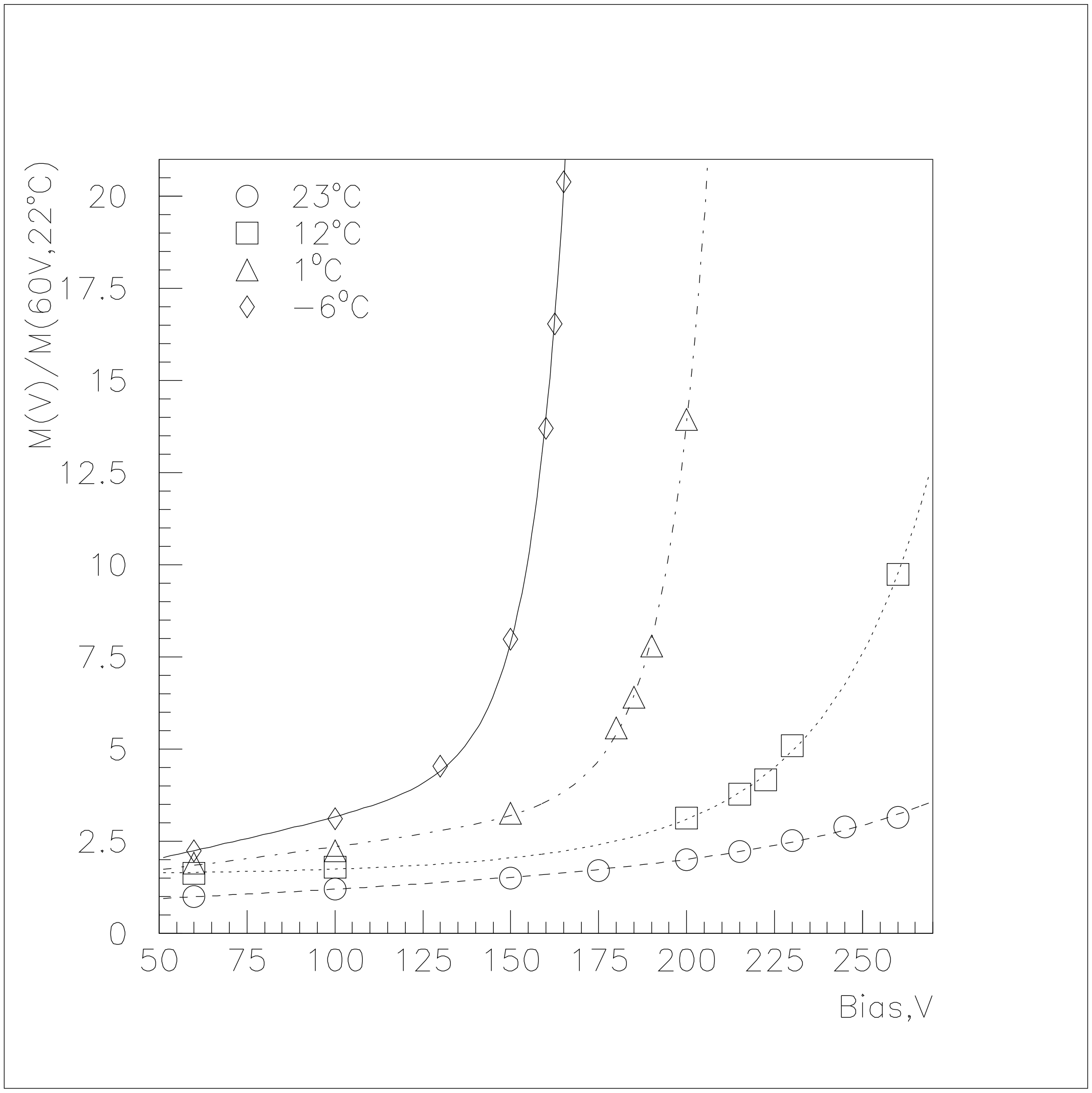

Fig. 5. The APD gain normalized to the gain at $60V$ and $22^oC$ as a function of the applied bias voltage at different temperatures.

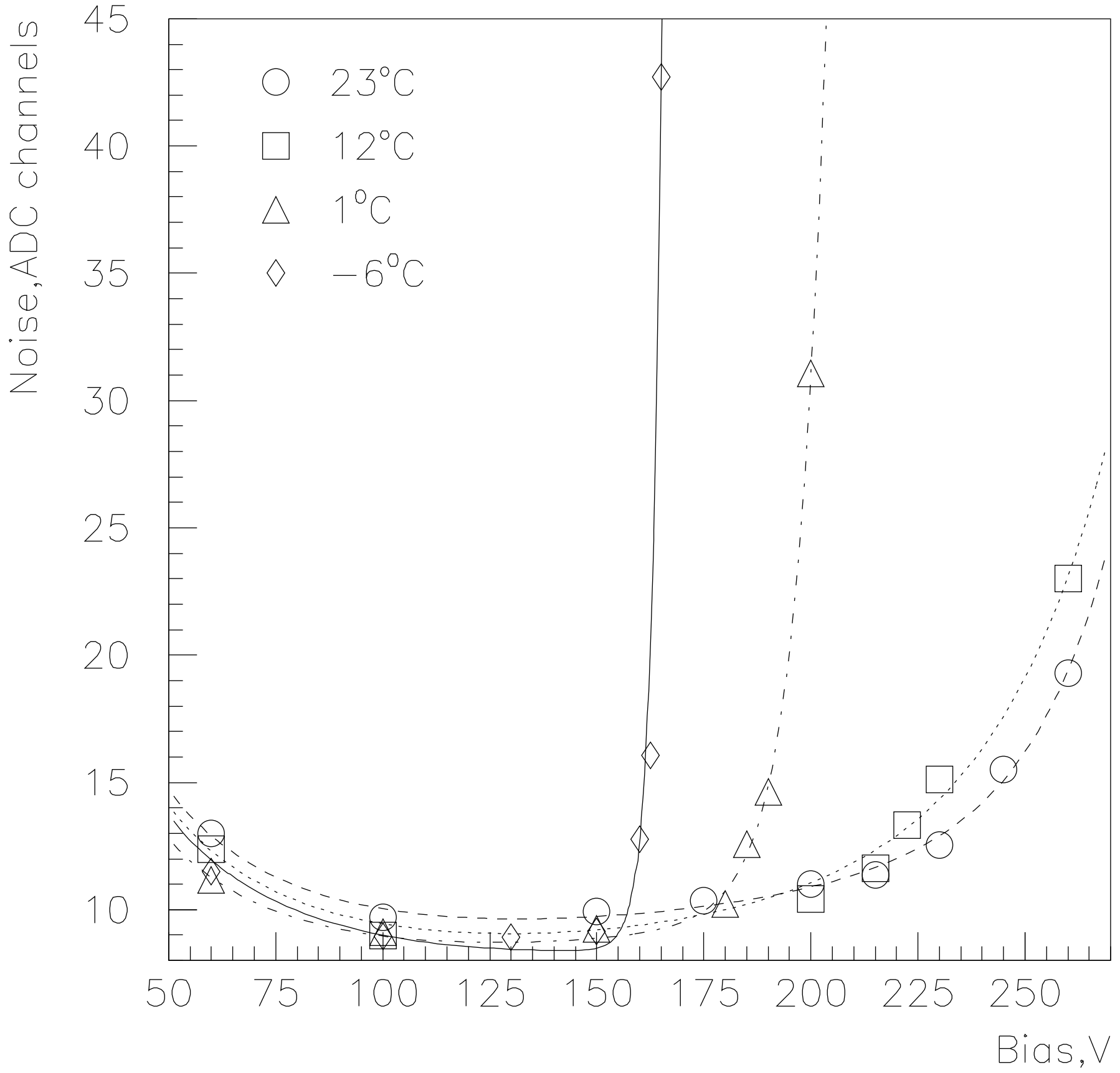

Fig. 6. The APD and preamplifier noise RMS as a function of the applied bias voltage at different temperatures.

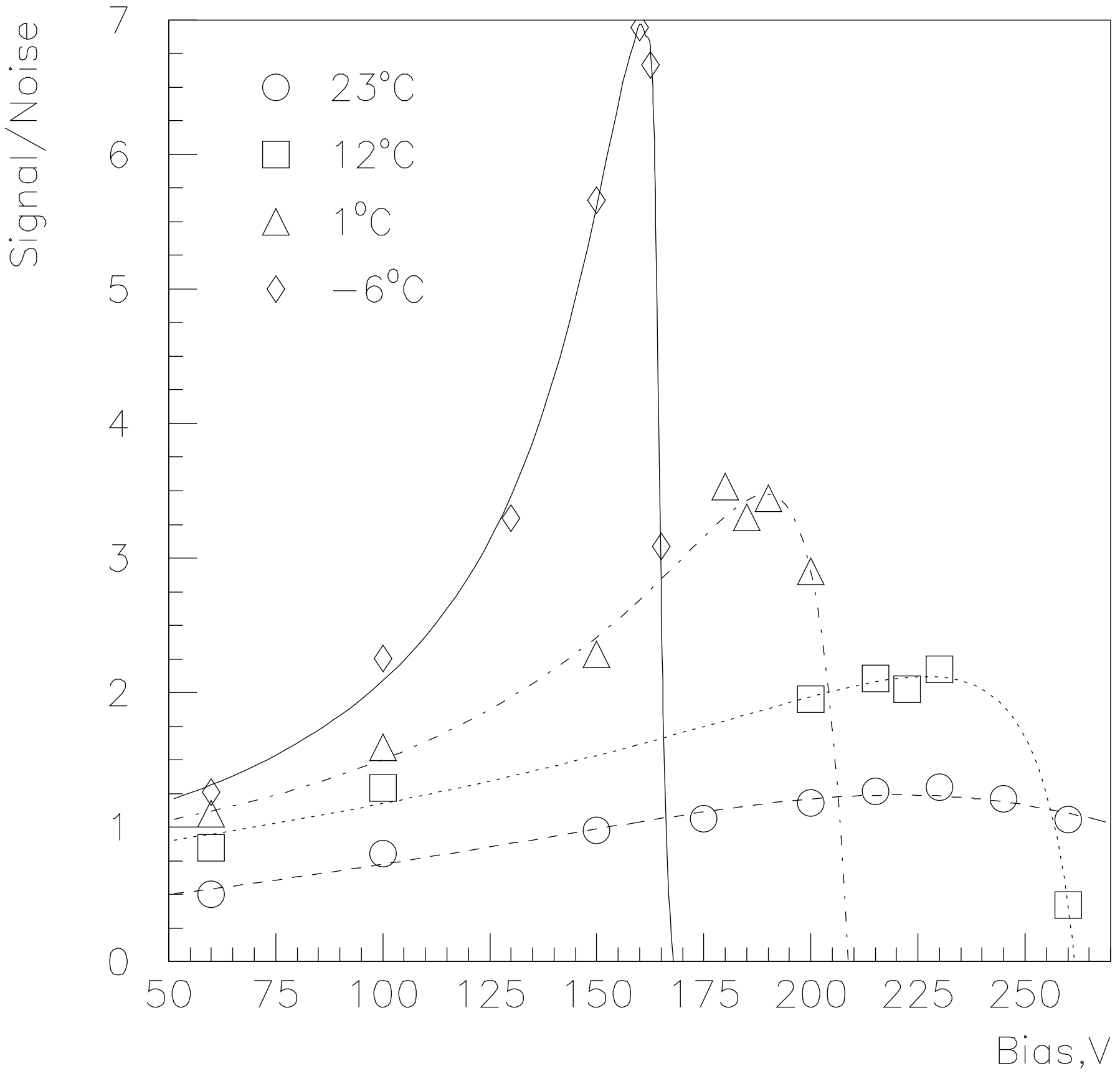

Fig. 7. The Signal/Noise ratio as a function of the applied bias voltage at different temperatures. LED intensity ≈ 35 photons.

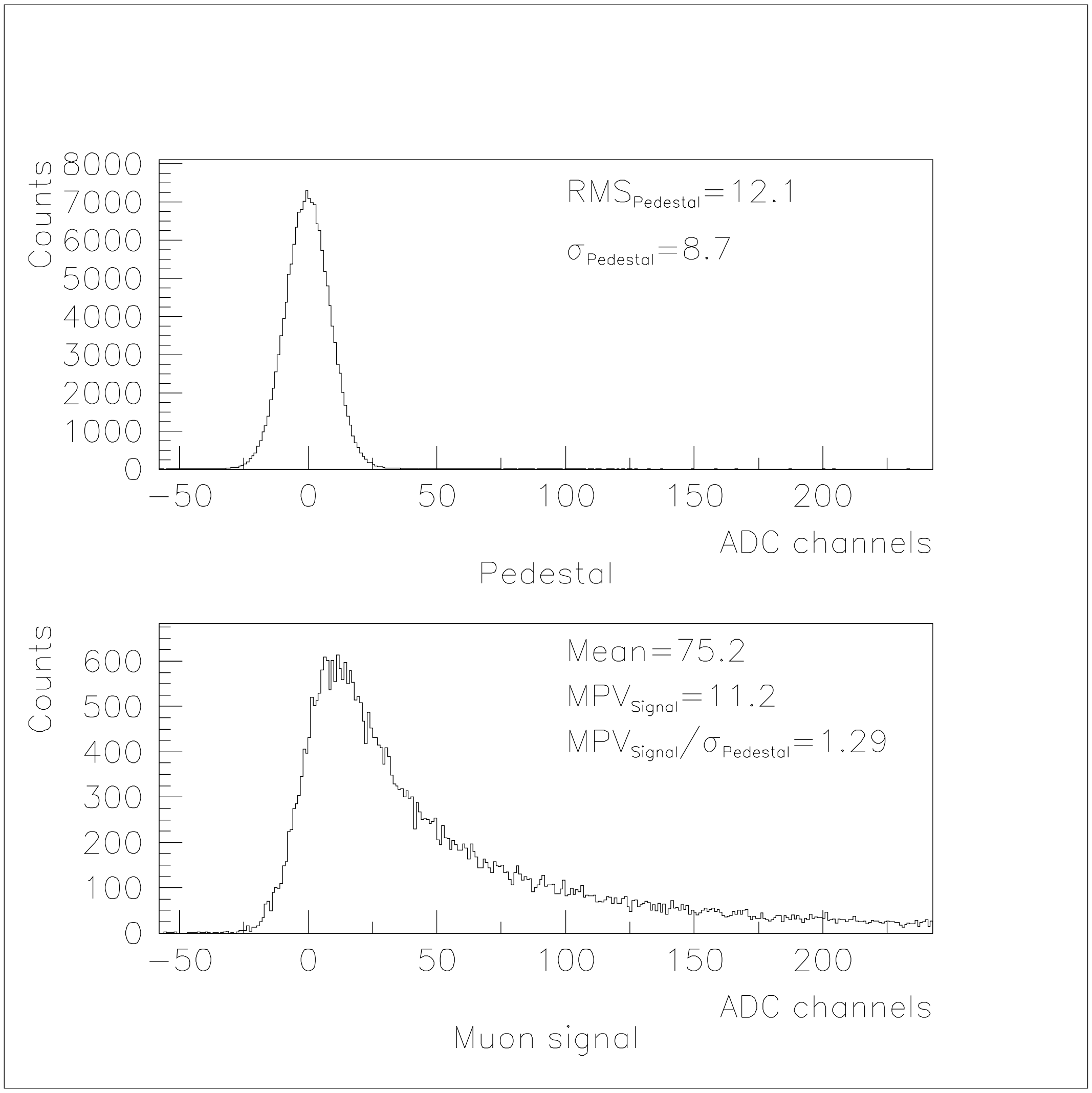

Fig. 8. Muon signal from $22 \times 22 \times 0.4cm^3$ scintillator with 2 coils of Bicron 99-172 fiber ($-8^oC$, $155V$ bias voltage, $160ns$ gate).

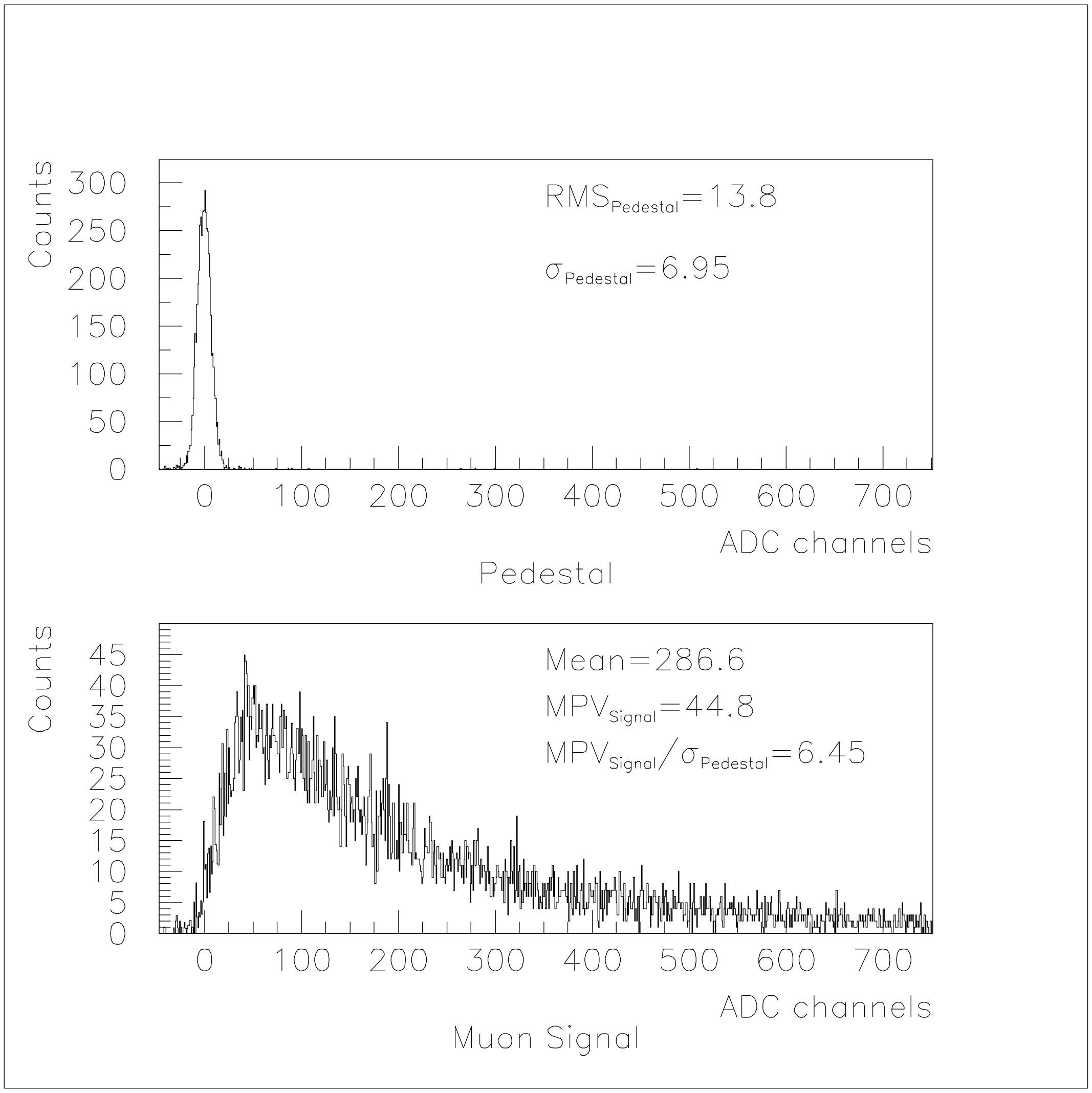

Fig. 9. Muon signal from $4 \times 4 \times 1cm^3$ scintillator with 10 coils of Bicron 99-172 fiber ($-9^oC$, $155V$ bias voltage, $160ns$ gate).

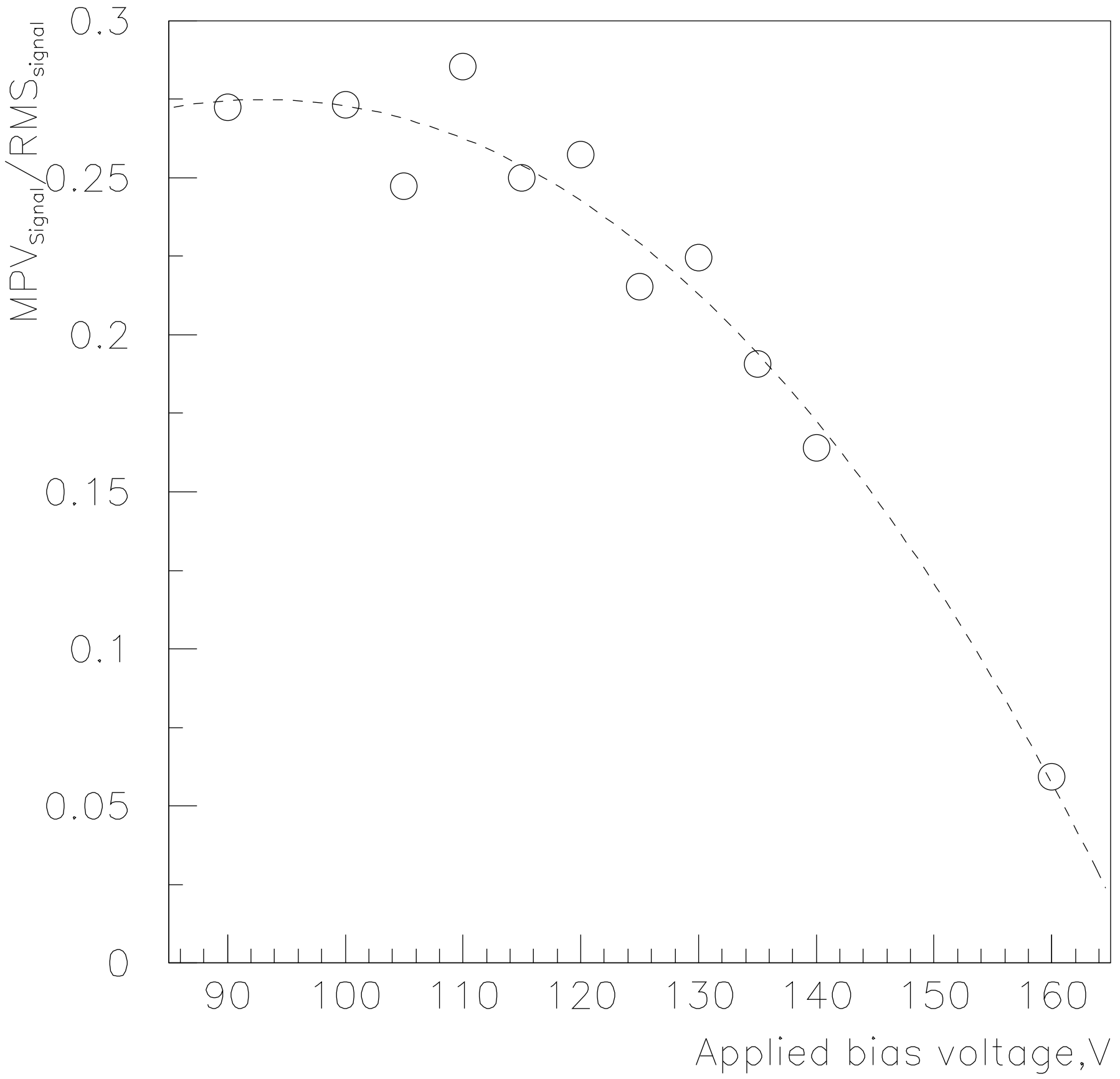

Fig. 10. The most probable value of the signal divided to $RMS$ of the signal as a function of the applied bias voltage. $4 \times 4 \times 1 cm^3$ scintillator with 10 coils of Bicron 99-172 fiber ($-9^o C$, $100 ns$ gate).

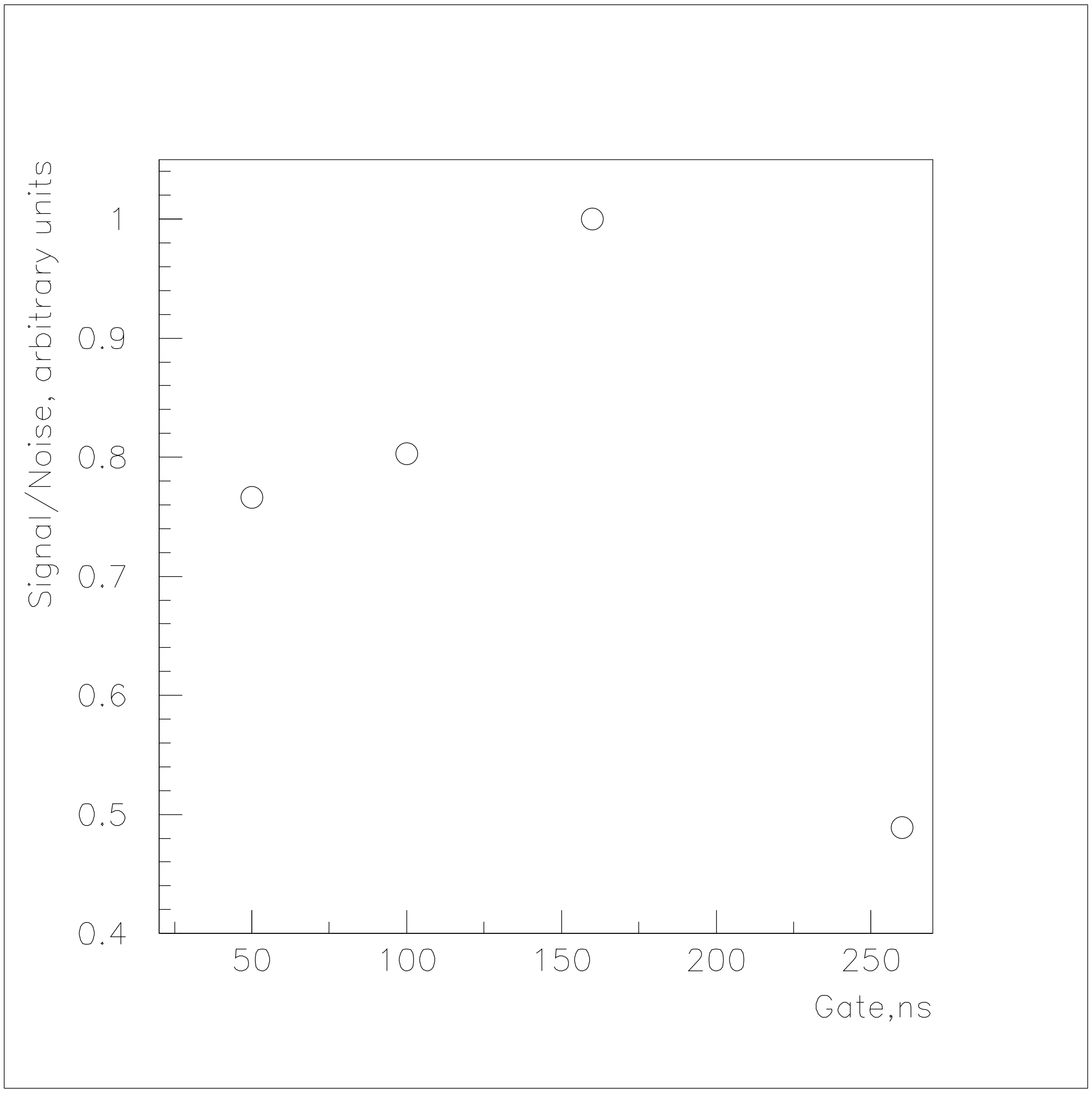

Fig. 11. The $Signal/Noise$ ratio as a function of the ADC gate. $4 \times 4 \times 1cm^3$ scintillator with 10 coils of Bicron 99-172 fiber ($-9^oC$, $153V$ bias voltage).

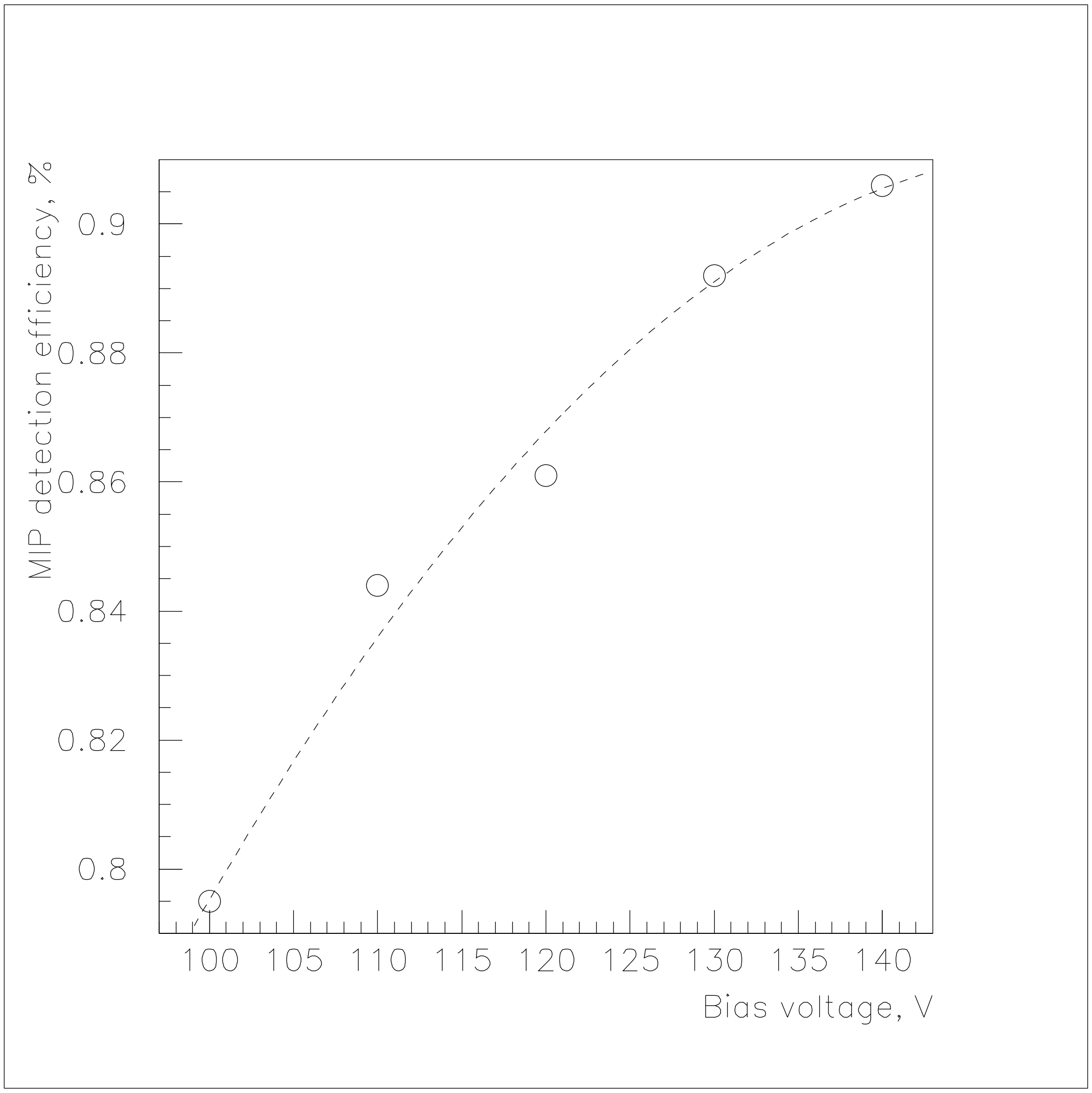

Fig. 12. The $MIP$ detection efficiency as a function of the applied bias voltage. $4 \times 4 \times 1cm^3$ scintillator with 10 coils of Bicron 99-172 fiber ($-9^oC$, $100ns$ gate).

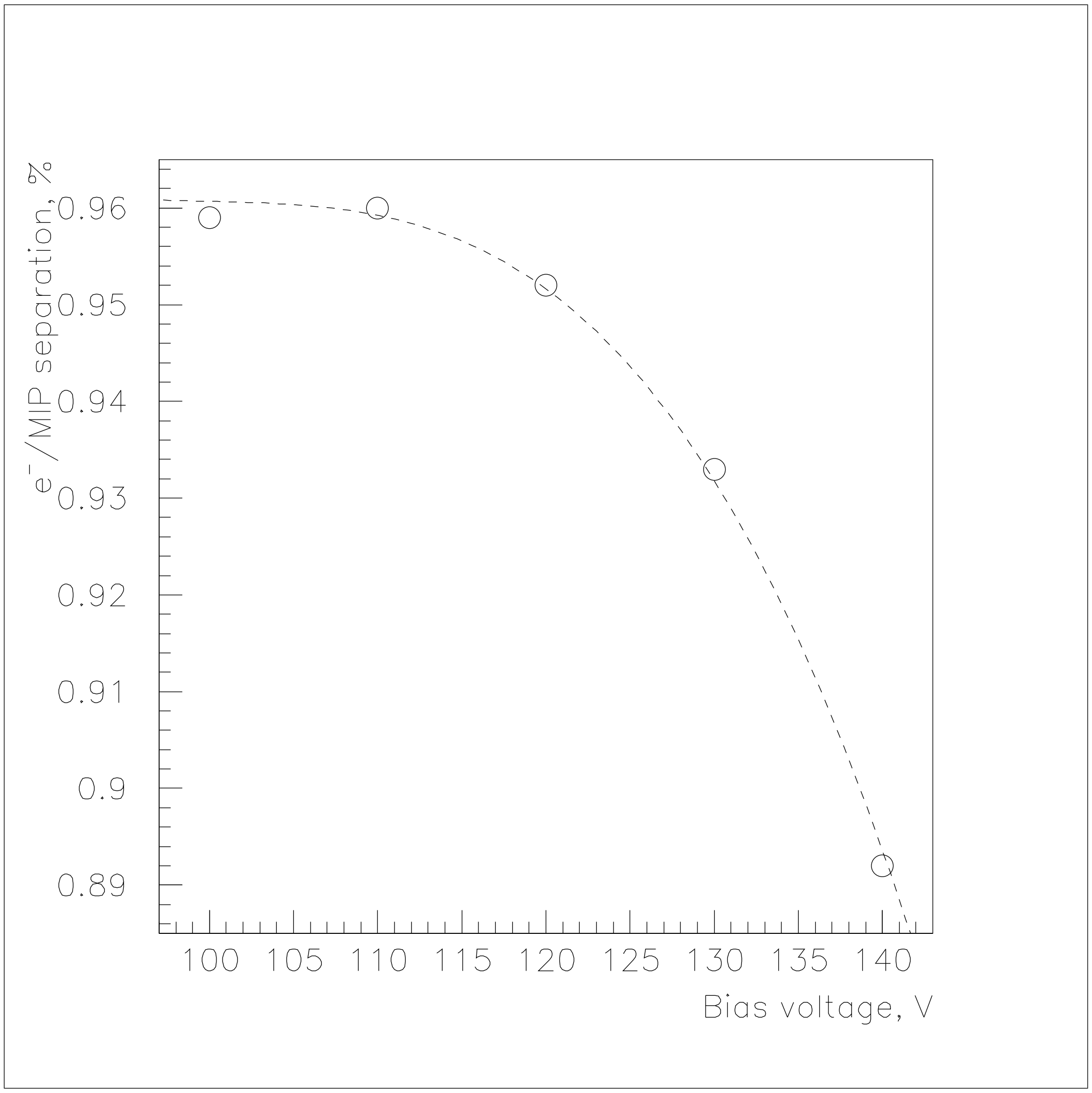

Fig. 13. The $e^{-}/MIP$ separation as a function of the applied bias voltage. $4 \times 4 \times 1cm^3$ scintillator with 10 coils of Bicron 99-172 fiber ($-9^oC$, $100ns$ gate).